\documentclass[twoside,12pt]{article}
\usepackage{epsfig,graphicx}
\usepackage{subfigure}
\usepackage[fleqn]{amsmath}
\usepackage{amssymb}
\usepackage{amsfonts}
\usepackage{mathrsfs}				
\usepackage{dsfont}					
\DeclareSymbolFont{rsfs}{U}{rsfs}{m}{n}
\DeclareSymbolFontAlphabet{\mathrsfs}{rsfs}

\allowdisplaybreaks[1] 

\def\Journal#1#2#3#4{{#1} {#2} (#4) #3 }

\def\NPA{{\em Nucl. Phys.} A}

\def\PLB{{\em Phys. Lett.} B}

\def\PRL{\em Phys. Rev. Lett.}

\def\PREP{\em Phys. Rep.}

\def\PRD{{\em Phys. Rev.} D}
\def\PRC{{\em Phys. Rev.} C}

\newcommand{\be}{\begin{equation}}
\newcommand{\ee}{\end{equation}}
\newcommand{\bea}{\begin{eqnarray}}
\newcommand{\eea}{\end{eqnarray}}

\newcommand{\D}[1]{\mathrm{d} \hspace{-1pt}#1}
\newcommand{\Dp}[2]{\mathrm{d}^{#1} \hspace{-1pt}#2}

\newcommand{\TrCD}{\text{Tr}_\text{C,D}}
\newcommand{\TrC}{\text{Tr}_\text{C}}
\newcommand{\TrD}{\text{Tr}_\text{D}}
\newcommand{\ld}{: \hspace{-0pt}}
\newcommand{\rd}{\hspace{-0pt} :}

\newcommand{\one}{\mathds{1}}

\newcommand{\sr}[1]{\shoveright{1}}

\newcommand{\GGV}{\langle : \hspace{-3pt} \frac{\alpha_s}{\pi} G^2 \hspace{-3pt} : \rangle}

\newcommand{\GGVren}{\langle \hspace{-0pt} \frac{\alpha_s}{\pi} G^2 \hspace{-0pt} \rangle}

\begin{document}

\title{ \vspace{1cm}Four-quark condensates in open-charm\\chiral QCD sum rules}

\author{{\sc T.\ Hilger,$^{1}$ T.\ Buchheim,$^{2}$ B.\ K\"ampfer,$^{1,2}$ S.\ Leupold$^3$}
\\
\it
$^1$Helmholtz-Zentrum Dresden Rossendorf, PF 510119, D-01314 Dresden,\\ \it Germany\\
\it
$^2$TU Dresden, Institut f\"ur Theoretische Physik, D-01062 Dresden,\\ \it Germany\\
\it
$^3$Institutionen f\"or fysik och astronomi, Uppsala Universitet,\\ \it SE-751 05 Uppsala, Sweden\\
}

\maketitle

\begin{abstract}
Recently, in \cite{Hilger:2011cq} QCD sum rules for chiral partners in the open-charm meson sector have been presented at nonzero baryon net density or temperature up to and including mass dimension 5.
Referring to this, details concerning the cancellation of infrared divergences are presented and important technical and conceptional ingredients for an incorporation of four-quark condensates beyond factorization and of other mass dimension 6 condensates are collected.
\end{abstract}


\section{Introduction}



A natural way to investigate dynamical chiral symmetry breaking is via considering the spectral difference of chiral partners.
In \cite{Hilger:2011cq}, chiral partner sum rules for heavy-light quark mesons in the spin 0 and spin 1 channels have been presented.
For heavy-light quark mesons, a crucial step in the evaluation of the operator product expansion is the cancellation of infrared divergences in virtue of the introduction of non-normal ordered condensates.
Thereby, the absorption of infrared divergent Wilson coefficients into the condensates at non-zero densities or temperatures requires additional renormalization relations.
In \cite{HilgerDA}, in-medium relations which are consistent to the vacuum case have been given.
However, only the pseudoscalar case has been investigated there and, moreover, a general proof to all orders in the mass dimension is still missing.
Chiral partner sum rules may, therefore, serve as a consistency check of the relations derived in \cite{HilgerDA}.

In view of the dynamical chiral symmetry breaking of quantum chromodynamics and order parameters of the symmetry restoration (phase) transition, QCD sum rules for heavy-light quark pseudoscalar mesons provide the amplification of the numerical impact of the chiral condensate in virtue of the heavy quark mass.
Hence, four-quark condensates, which are of mass dimension 6, are numerically not as important as in the case of mesons consisting of light quarks only.
However, in order to determine their numerical impact one has to take them into account.
The evaluation of Wilson coefficients for four-quark condensates in case of heavy-light quark sum rules is an elaborate task.
A common approximation is the vacuum saturation hypothesis, which leads to a factorization of four-quark condensates into squares of the chiral condensate.
Once applied, the factorization significantly simplifies the calculation.
However, as only certain combinations of four-quark condensates may serve as order parameters of chiral symmetry the factorization of all four-quark condensates in terms of the chiral condensate is not necessarily consistent.\footnote{Note, that the factorization of four-quark condensates into squares of the chiral condensate is exact in the limit of infinitely many colors.}
Therefore, the extension of mass dimension 5 in-medium heavy-light quark QCD sum rules to mass dimension 6 requires the consistent evaluation of four-quark condensate contributions.
With respect of this task, chiral partner sum rules provide assistance as only chirally odd condensates do not cancel out and the number of terms which have to be calculated is significantly reduced.

%
%

\section{Chiral partner heavy-light quark meson correlators}
\label{sct:correlators}

We consider the currents
\begin{subequations}
\label{eq:def_current_2_flavor}
\begin{align}
j^{{\rm S}}(x) & :=  \bar q_1(x) \, q_2(x)  \: , 
&
j^{{\rm P}}(x) & :=  \bar q_1(x) \, i\gamma_5 \, q_2(x) \: , \label{eq:def_pscalar_2_flavor} \\
j_\mu^{{\rm V}}(x) & :=  \bar q_1(x) \, \gamma_\mu \, q_2(x) \: , 
&
j_\mu^{{\rm A}}(x) & :=  \bar q_1(x) \, \gamma_5 \gamma_\mu \, q_2(x) \label{eq:def_avector_2_flavor}
\end{align}
\end{subequations}
and the corresponding causal correlators
\begin{subequations}\label{eq:def_corr_2_flavor}
\begin{align}
	\label{eq:corrdef_2_flavor_0}
	\Pi^{\rm (S,P)}(q) &= i \int \Dp{4}{x} e^{iqx} \langle {\rm T} \left[ j^{\rm (S,P)}(x) j^{\rm (S,P)\dagger}(0) \right] \rangle
	\: ,
	\\
	\label{eq:def_corr_2_flavor_1}
	\Pi^{\rm (V,A)}_{\mu\nu}(q) &= i \int \Dp{4}{x} e^{iqx} \langle {\rm T} \left[ j^{\rm (V,A)}_\mu(x) j^{\rm (V,A)\dagger}_\nu(0) \right] \rangle
	\: ,
\end{align}
\end{subequations}
where T$[\ldots]$ denotes time-ordering and $\langle \ldots \rangle$ means Gibbs averaging \cite{Bochkarev:1985ex}.
In the rest frame of the nuclear medium, i.e.\ $n=(1,\vec 0)$, and for mesons at rest, i.e.\ $q=(1,\vec 0)$, the (axial-) vector correlator can be decomposed as
\begin{equation}
\label{eq:proj_VA}
	\Pi^{\rm(V,A)}_{\mu\nu}(q) = \left( \frac{q_\mu q_\nu}{q^2} - g_{\mu\nu}\right) \Pi^{\rm(V,A)}_{\rm T}(q)
		+ \frac{q_\mu q_\nu}{q^2} \Pi^{\rm(V,A)}_{\rm L}(q)
\end{equation}
with
$
	\Pi^{\rm(V,A)}_{\rm T}(q) = \frac{1}{3} \left( \frac{q^\mu q^\nu}{q^2} - g^{\mu\nu} \right) \Pi^{\rm(V,A)}_{\mu\nu}(q)
$ and 
$
	\Pi^{\rm(V,A)}_{\rm L}(q) = \frac{1}{q^2} q^\mu q^\nu \Pi^{\rm(V,A)}_{\mu\nu}(q)
$ written manifestly covariant.
The transversal projection for a vanishing light quark mass is given by
\begin{equation}
\label{eq:vaps}
    \Pi^{\rm V-A}_{\rm T}(q) = - \frac{m_2^2}{3q^2} \Pi^{\rm P-S}(q) - \frac{1}{3} \Pi^{{\rm V-A}}(q) - \frac23 \frac{m_2}{q^2} \langle \bar{q}_1 q_1 \rangle
    \: ,
\end{equation}
with $\Pi^{{\rm (V,A)}}(q) = g^{\mu\nu} \Pi^{{\rm (V,A)}}_{\mu\nu}(q)$.

\section{Cancellation of IR divergences}
\label{sct:chiral_partner_IR_cancellation}

In lowest-order perturbation theory the operator product expansion within the background field method \cite{Shifman:1978bx,Novikov:1983gd} for the difference of chiral partner correlators reads
\begin{subequations} \label{eq:3}
\begin{multline}
	\label{eq:3a}
    \Pi^{{\rm P-S}(0)}(q) = -i \int \frac{\Dp{4}{p}}{(2\pi)^4} \langle \ld \frac12 \TrC \left\{
        \TrD[S_2(p+q)] \TrD[S_1(p)]
        \right.
        \\
        + \frac12 \TrD[S_2(p+q)\sigma_{\mu\nu}] \TrD[S_1(p) \sigma^{\mu\nu}]
        \left.
        + \TrD[S_2(p+q) \gamma_5] \TrD[S_1(p)\gamma_5]
        \right\} \rd \rangle \: ,
\end{multline}
\begin{multline}
    \label{eq:3b}
    \Pi^{{\rm P-S}(2)}(q) = \sum_n \frac{(-i)^n}{n!} \frac12 \sum_\Gamma^{\{\mathds{1}, \sigma_{\alpha < \beta}, \gamma_5 \} }
        \langle \ld
        \overline{q}_1 \overleftarrow{D}_{\vec\alpha_n}
        \Gamma
        \partial^{\vec\alpha_n}
        \left( \TrD[ \Gamma S_2(q)] \right)
        q_1
        \\
        + \overline{q}_2 \Gamma \partial^{\vec\alpha_n}
        \left( \TrD[ \Gamma S_1(-q)] \right)
        \overrightarrow{D}_{\vec\alpha_n} q_2
        \rd \rangle
\end{multline}
\begin{multline}
    \label{eq:3c}
    \Pi^{{\rm V-A}(0)}(q) = i \int \frac{\Dp{4}{p}}{(2\pi)^4} \langle \ld 2 \TrC \left\{
        \TrD[S_2(p+q)] \TrD[S_1(p)]
        \right.
        \\
        \left.
        - \TrD[S_2(p+q) \gamma_5] \TrD[S_1(p)\gamma_5]
        \right\} \rd \rangle \: ,
\end{multline}
\begin{multline}
    \label{eq:3d}
    \Pi^{{\rm V-A}(2)}(q) = - \sum_n \frac{(-i)^n}{n!} 2 \sum_\Gamma^{\{\mathds{1},i\gamma_5\}}
        \langle \ld
        \overline{q}_1 \overleftarrow{D}_{\vec\alpha_n}
        \Gamma
        \partial^{\vec\alpha_n}
        \left( \TrD[ \Gamma S_2(q)] \right)
        q_1
        \\
        + \overline{q}_2 \Gamma \partial^{\vec\alpha_n}
        \left( \TrD[ \Gamma S_1(-q)] \right)
        \overrightarrow{D}_{\vec\alpha_n} q_2
        \rd \rangle \: ,
\end{multline}
\end{subequations}
where $\TrCD$ denotes trace w.r.t.\ color or Dirac indices.
$D_\mu(x) = \partial_\mu - i g \mathrsfs{A}_\mu(x)$ is the covariant derivative and an arrow indicates whether it acts to the left or right.
For sake of a concise notation we have defined ${D}_{\vec{\alpha}_n} = D_{\alpha_1} \ldots D_{\alpha_n}$ (with an analog notation for the partial derivative).
The quark propagator is denoted by
$iS_{ij}(p) = \int \Dp{4}{x} e^{ipx} \langle \Omega | {\rm T} \left[ q_i(x) \bar q_j(0) \right] | \Omega \rangle$,
$| \Omega \rangle$ is the ground state of the strong interaction,
$\sigma_{\mu\nu} = i\left[\gamma_\mu,\gamma_\nu\right]/2$
and $: \ldots :$ means normal ordering w.r.t.\ the perturbative vacuum $| 0 \rangle$.
Eq.\ \eqref{eq:3} stems from the application of Wicks theorem to \eqref{eq:def_corr_2_flavor} in lowest order of the perturbative expansion.
$\Pi = \Pi^{(0)}+\Pi^{(2)}$, where $\Pi^{(0)}$ corresponds to the term where the quark fields of the currents \eqref{eq:def_current_2_flavor} are contracted to propagators and $\Pi^{(2)}$ to the term with 2 not contracted quark fields.
There is no flow of hard momentum in the term with four not contracted quark fields.
Hence, it does not contribute.

In \cite{Hilger:2011cq}, it has been proven that $\TrD[S_1(p) \Gamma] = 0$ for $m_1 = 0$, $p^2 \neq 0$ and $\Gamma \in \{\mathds{1}, \sigma_{\mu\nu}, \gamma_5\}$. 
Hence, a superficial view on \eqref{eq:3} may tempt to the conclusion that $\Pi^{(0)}$ is zero for chiral partner operator product expansions of heavy-light quark meson currents.
On the other hand it is clear that only matrix elements of chirally odd operators may enter the OPE, whereas gluon condensates are chirally even.
In this sense, the cancellation of $\Pi^{(0)}$ is in line with naive expectations.
But the introduction of non-normal ordered condensates via $\Pi^{(2)}$ would introduce gluon condensates together with infrared divergences.
Taking this as a heuristic argument for non-zero $\Pi^{(0)}$ raises the question of the precise cancellation of these terms and terms added by introducing non-normal ordered condensates.
Indeed, from the in-medium OPE of D mesons up to mass dimension 5 \cite{HilgerDA,Hilger:2010zb} it is known that the medium specific divergences are canceled because of the renormalization of $\langle \ld \bar d \gamma_\mu D_\nu d \rd \rangle$.
But this term is chirally even (formally, it is the matrix element of a vector current) and does not enter \eqref{eq:3}.
Moreover, it is known that introducing non-normal ordered condensates, in order to cancel infrared divergent Wilson coefficients of gluon condensates, leads to additional finite gluon contributions.
Clearly, these have to cancel out in case of chiral partner sum rules.
Again, this can be taken as a heuristic argument that only those mass divergences can remain in $\Pi^{(0)}$ which are cancelled by chirally odd condensates.
Two questions have to be answered.
Do all infrared divergences cancel out? And does the renormalization procedure introduce chirally even condensates?
Thus, a careful analysis is mandatory to prove that the obtained results are infrared stable and that the renormalization procedure is consistent.

The quark propagator in fixed-point gauge $(x-x_0)_\mu \mathrsfs{A}^\mu(x) = 0$ \cite{Novikov:1983gd} can be written as
$
S(p) = \sum_{n=0}^\infty S^{(n)}(p)
$
with
$
    S^{(n)}(p)
        = (-1) S^{(n-1)}(p) \left( \gamma \tilde{A} \right) S^{(0)}(p)
        = (-1) S^{(0)}(p) \left( \gamma \tilde{A} \right) S^{(n-1)}(p) \: .
$
$\tilde{A}$ denotes a derivative operator which arises due to the Fourier transform 
of the perturbation series for the quark propagator in coordinate space 
from the gluonic background field $\mathrsfs{A}_{\mu}$;
$
    \tilde{A}_{\mu} = \sum_{n=0}^\infty \tilde{A}_{\mu}^{(n)}
$
with
$
    \tilde{A}_{\mu}^{(n)} = - \frac{(-i)^{n+1} g}{n! (n+2)} \left(
        D_{\alpha_1} \ldots D_{\alpha_n} \mathrsfs{G}_{\mu\nu}(0) \right)
        \partial^{\nu} \partial^{\alpha_1} \ldots \partial^{\alpha_n},
$
where $\mathrsfs{G}_{\mu\nu} = i \left[ D_\mu, D_\nu \right]/g = G_{\mu\nu}^A t^A$ is the gluon field strength tensor and $g=\sqrt{4\pi \alpha_s}$ is the coupling.
$t^A$ are the generators of the color group and $A=1, \ldots, N_c^2 -1$.
For the contributions to \eqref{eq:3a} up to mass dimension 5, which restricts the quark propagator to next-to-next-to leading order and the gluon field to lowest order, the following traces are evaluated
\begin{subequations}
\begin{flalign}
	\TrD\left[ S^{(0)}(p)\right] &= \frac{4 m}{p^2-m^2} \: ,
	&
	\TrD\left[ S^{(0)}(p) \gamma_5 \sigma_{\mu\nu}\right] &= 0 \: ,
	&
	\TrD\left[ S^{(0)}(p) \gamma_5 \right] &= 0 \: ,
	\label{eq:S0_gamma_5_trace}
	\\
	\TrD\left[ S^{(1)}(p) \gamma_5 \sigma_{\mu\nu} \right] &= -i g \frac{2m}{(p^2-m^2)^2} \epsilon_{\mu\nu\kappa\lambda} \mathrsfs{G}^{\kappa\lambda} \: ,
	&
	\TrD\left[ S^{(1)}(p)\right] &= 0 \: ,
	&
	\TrD\left[ S^{(1)}(p) \gamma_5\right] &= 0 \: ,
	\\
	\TrD\left[ S^{(2)}(p)\right] &= 8 g^2 \frac{m p_\mu p_\alpha}{(p^2-m^2)^4} \mathrsfs{G}^{\mu\nu} \mathrsfs{G}^{\alpha}_{\phantom{\alpha}\nu} \: .
\end{flalign}
\end{subequations}
It is not necessary to consider traces of the second order quark propagator with $\gamma_5$ and $\sigma_{\mu\nu}$.
Up to mass dimension 5 they can only be multiplied with their lowest order counterparts \eqref{eq:S0_gamma_5_trace}, which are zero.
A combination of second-order and first-order propagator leads to mass dimension 6 terms.
Using
$\TrD[S_1 \gamma_5 \sigma_{\mu\nu}] \TrD\left[S_2 \gamma_5 \sigma^{\mu\nu}\right]
		= \TrD\left[S_1 \sigma_{\mu\nu}\right] \TrD\left[S_2 \sigma^{\mu\nu}\right]
$
the following contributions have to be considered
\begin{multline} \label{eq:chiral_partner_div}
	    \Pi^{{\rm P-S}(0)}(q)
	    \\
	    = -i \int \frac{\Dp{4}{p}}{(2\pi)^4} \langle \ld \frac12 \TrC \biggl\{
        \TrD[S^{(0)}_c(p+q)] \TrD[S^{(0)}_d(p)]
        + \TrD[S^{(2)}_c(p+q)] \TrD[S^{(0)}_d(p)]
        \biggr.
        \\ \left.
        + \TrD[S^{(0)}_c(p+q)] \TrD[S^{(2)}_d(p)]
        + \frac12 \TrD[S^{(1)}_c(p+q)\sigma_{\mu\nu}] \TrD[S^{(1)}_d(p) \sigma^{\mu\nu}]
        \right\} \rd \rangle \: .
\end{multline}
Due to $\TrD [S^{(0)} \gamma_5] = \TrD [S^{(1)} \gamma_5] = 0$ for arbitrary quark masses, there is no $\gamma_5$ contribution up to this mass dimension.
Likewise $\TrD [S^{(0)} \gamma_5 \sigma_{\mu\nu}] = \TrD [S^{(1)}] = 0$.
Lorentz invariance requires that terms with only one gluon field are zero.
Therefore, a first-order quark propagator must be combined with a propagator of at least the same order.
Keeping both quark-masses $m_{d,c}$ finite these four terms give rise to the following integrals
\begin{multline} \label{eq:chiral_partner_div2}
	    \Pi^{{\rm P-S}(0)}(q)
	    \\
	    = -i \int \frac{\Dp{4}{p}}{(2\pi)^4} \langle \ld
        \frac 12 \frac{4 m_c}{(p+q)^2-m_c^2} \frac{4 m_d}{p^2-m_d^2} \TrC \left[\one_{\rm C}\right]
        \\
        + \frac {g^2}2 \frac{8 m_c}{[(p+q)^2-m_c^2]^4} \frac{4 m_d}{p^2-m_d^2} p^\mu p^\alpha G^A_{\mu\nu} G^{B \phantom{\mu}\nu}_{\phantom{B} \alpha}
        	\TrC \left[t^A t^B\right]
        \\
        + \frac {g^2}{2} \frac{4 m_c}{(p+q)^2-m_c^2} \frac{8 m_d}{[p^2-m_d^2]^4} p^\mu p^\alpha G^A_{\mu\nu} G^{B \phantom{\alpha}\nu}_{\phantom{B} \alpha}
        	\TrC \left[t^A t^B\right]
        \\
        + \frac{ g^2}{4} \frac{4 m_c}{[(p+q)^2-m_c^2]^2} \frac{4 m_d}{[p^2-m_d^2]^2} G^A_{\mu\nu} G^{B \mu\nu}
        	\TrC \left[t^A t^B\right]
        \rd \rangle \: .
\end{multline}
Analyzing the integrals in euclidean space in terms of the integral
\begin{equation}
	{\rm I}_{ijk}(q^2,m_d^2,m_c^2)
	= \int_0^1 \D{\alpha} \frac{\alpha^i (1-\alpha)^j}{(\alpha(1-\alpha)q^2 + \alpha m_d^2 + (1-\alpha) m_c^2)^k}
\end{equation}
reveals the following results.
The first term is $\propto m_d m_c {\rm I}_{0,0,0}(q^2,m_d^2,m_c^2)$.
The second term generates two terms which are $\propto m_d m_c {\rm I}_{0,3,2}(q^2,m_d^2,m_c^2)$ and $\propto m_d m_c {\rm I}_{2,3,3}(q^2,m_d^2,m_c^2)$, respectively.
The last term is $\propto m_d m_c {\rm I}_{1,1,2}(q^2,m_d^2,m_c^2)$.
These terms are all zero in the limit $m_d \to 0$.
On the other hand, the third term does not vanish for $m_d \to 0$.
It causes a term $\propto m_d m_c {\rm I}_{3,0,2}(q^2,m_d^2,m_c^2)$, which diverges with $m_d^{-1}$.
Projection of the Lorentz indices leads to
\begin{equation} \label{eq:special_gluon_projection}
	\langle \ld \delta^{AB} G^A_{\mu\nu} G^{B \phantom{\alpha}\nu}_{\phantom{B} \alpha} \rd \rangle
	= \frac{g_{\mu\alpha}}{4} \langle \ld G^2 \rd \rangle
	- \frac 13 \left( g_{\mu\alpha} - 4 \frac{v_\mu v_\alpha}{v^2} \right)
	\langle \ld \left( \frac{ \left(vG\right)^2}{v^2} - \frac{G^2}{4} \right) \rd \rangle
	\: ,
\end{equation}
where $v_\mu$ is the medium four velocity, $\left(vG\right)^2 = v^\mu v^\nu g^{\alpha\beta} G^A_{\mu\alpha} G^A_{\nu\beta}$ and $G^2 = G^A_{\mu\nu} G^{A\nu\beta}$.
Note that the diagonal elements of \eqref{eq:special_gluon_projection} are vacuum specific, whereas the medium specific contribution is traceless.
The integral of the third term of \eqref{eq:chiral_partner_div} can be evaluated in euclidean space:
\begin{multline}
\label{eq:IR_divergences_critical_integral}
	\int \frac{\Dp{4}{p}}{(2\pi)^4} \frac{m_d m_c}{[(p-q)^2 - m_d^2]^4} \frac{(p-q)_\mu (p-q)_\alpha}{p^2-m_c^2}
		\\
		\stackrel{\rm W.R.}{\longrightarrow} \frac 23 \frac{m_c m_d}{(4\pi)^2}
			\left( \frac{g_{\mu\alpha}}{2} {\rm I}_{3,0,2}(q^2,m_d^2,m_c^2) + 2 q_\mu q_\alpha {\rm I}_{3,2,3}(q^2,m_d^2,m_c^2) \right)
			\: ,
\end{multline}
where
${\rm I}_{5,0,3} - 2 {\rm I}_{4,0,3} + {\rm I}_{3,0,3} = {\rm I}_{3,2,3}$
has been used.
By virtue of ${\rm I}_{ijk}(q^2,m_d^2,m_c^2) \to {\rm I}_{jik}(q^2,m_d^2,m_c^2)$ the corresponding integral of the second term in \eqref{eq:chiral_partner_div2} can be derived.
The limit $m_d \to 0$ for both terms in \eqref{eq:IR_divergences_critical_integral} is
\begin{subequations}\label{eq:IR_divergences_limits}
\begin{align}
	\label{eq:IR_divergences_limits_a}
	\lim_{m_d \to 0} m_d {\rm I}_{3,0,2}(q^2,m_d^2,m_c^2) & = \frac{1}{m_d} \frac{1}{q^2+m_c^2}
	\: ,
	\\
	\lim_{m_d \to 0} m_d {\rm I}_{3,2,3}(q^2,m_d^2,m_c^2) & = 0
	\: ,
\end{align}
\end{subequations}
where \eqref{eq:IR_divergences_limits_a} reveals the famous infrared singularity from vacuum D meson sum rules.
As the medium specific contribution to \eqref{eq:special_gluon_projection} is traceless with respect to Lorentz indices and due to the vanishing of the second term in \eqref{eq:IR_divergences_critical_integral} there is no medium specific infrared divergent term.
Hence, the only terms that have to be absorbed into the condensates by virtue of the introduction of non-normal ordered condensates are vacuum specific.
This is in line with the cancellation of $\langle \ld \bar d \gamma_\mu D_\nu d \rd \rangle$ in \eqref{eq:3} which would have to absorb the medium specific divergences.
Moreover, infrared divergent terms which enter through the medium specific gluon condensate, therefore, must enter the $\gamma_\mu$ or $\gamma_5 \gamma_\mu$ parts of the quark propagator.

Owing to \eqref{eq:special_gluon_projection}, \eqref{eq:IR_divergences_critical_integral} and \eqref{eq:IR_divergences_limits} the limit $m_d \to 0$ of \eqref{eq:chiral_partner_div} in euclidean space is
\begin{multline}
	\Pi^{{\rm P-S}(0)}(q)
		\\
		= - \frac i2 \int \frac{\Dp{4}{p}}{(2\pi)^4} \langle \ld
        \TrD[S^{(0)}_c(p+q)] \TrCD[S^{(2)}_d(p)]
        \rd \rangle
        \stackrel{\rm W.R.}{\longrightarrow}
        - \frac i6 i\GGV {\rm I}_{3,0,2}(q^2,m_d^2,m_c^2) m_c m_d
        \\ \shoveright{
        = - \frac i6 i\GGV \frac{m_c}{m_d} \frac{1}{q^2+m_c^2}
        \: ,}
\end{multline}
where it has been used that the lowest order quark propagator is a unit in color space and the additional imaginary unit stems from the Wick rotation (W.R.).

The normal ordered chiral condensate in lowest order of the light quark mass enters \eqref{eq:3b} via the following expression \cite{Hilger:2011cq}
\begin{equation}
	\Pi^{{\rm P-S}(2)}_{\langle \bar d d \rangle}(q) = \frac 12 \langle \ld \bar d d \rd \rangle \TrD \left[ S_c(q) \right]
	\: .
\end{equation}
Here, $\TrD \left[ S_c(q) \right]$ is the Wilson coefficient of the chiral condensate.
Expressing the normal ordered condensate by the non-normal ordered condensate
\begin{equation}
\label{eq:chiral_cond_ren}
	\langle \ld \bar d d \rd \rangle = \langle \bar d d \rangle
		+ i \int \frac{\Dp{4}{p}}{(2\pi)^4} \langle \ld \TrCD \left[ S_d(p) \right] \rd \rangle
\end{equation}
leads to
\begin{equation} \label{eq:chiral_condensate_renormalization_contribution}
	\Pi^{{\rm P-S}(2)}_{\langle \bar d d \rangle}(q) = \frac 12 \langle \bar d d \rangle \TrD \left[ S_c(q) \right]
		+ \frac 12 \TrD \left[ S_c(q) \right] i \int \frac{\Dp{4}{p}}{(2\pi)^4} \langle \ld \TrCD \left[ S_d(p) \right] \rd \rangle
	\: .
\end{equation}
Despite the striking similarity, revealed by the projection onto elements of the Clifford basis in \eqref{eq:3}, of the third term in \eqref{eq:chiral_partner_div} which has to be canceled and the additional term in \eqref{eq:chiral_condensate_renormalization_contribution}, a general proof cannot be given and the necessity for a precise evaluation is obvious.
Evaluation of \eqref{eq:chiral_cond_ren} up to order $\alpha_s^1$ in $\overline{\rm MS}$-scheme gives
\begin{equation}
\langle\bar{q}q\rangle = \langle:\bar{q}q:\rangle
		+ \frac{3}{4\pi^2}m_q^3 \left( \ln{\frac{\mu^2}{m_q^2}}+1 \right)
		- \frac{1}{12m_q} \GGVren
		\label{physical_condensates_2a}
\end{equation}
with the renormalization scale $\mu$.
Insertion into \eqref{eq:chiral_condensate_renormalization_contribution} and in virtue of the limit $m_d \to 0$ the anticipated cancellation of infrared divergences in terms of light quark masses is revealed
\begin{equation} \label{eq:chiral_condensate_renormalization_contribution2}
	\Pi^{{\rm P-S}(2)}_{\langle \bar d d \rangle}(q)
	\stackrel{\rm W.R.}{\longrightarrow}
		\frac 12 \langle \bar d d \rangle \TrD \left[ S_c(q) \right]
		- \frac 16 \frac{m_c}{m_d} \frac{1}{q^2+m_c^2} \GGV
	\: .
\end{equation}
Adding \eqref{eq:IR_divergences_limits} and \eqref{eq:chiral_condensate_renormalization_contribution2} the infrared divergent term cancels out.
Furthermore, Eq.\ \eqref{eq:3} and the explicit evaluation of the renormalization of normal ordered condensates \cite{HilgerDA} shows, that in chiral partner sum rules up to and including mass dimension 5 in the limit $m_d \to 0$ only the chiral condensate mixes with other condensates by virtue of introducing non-normal ordered condensates.
Thus, apart from the term which cancels the infrared divergence no additional chirally even terms enter and the final operator product expansion is chirally odd.

So far, the investigation was carried out for the spin 0 case.
Fortunately, as the only formal difference between the spin 0 and spin 1 case is the cancellation of the projections of quark propagators onto the tensor $\sigma_{\mu\nu}$ in the spin 1 case and a global factor $-4$, see \eqref{eq:3}, from the previous evaluation it is clear that the terms of interest are the same in both cases up to mass dimension 5.

\section{Comments about four-quark condensates}

\begin{figure}
\centering
\subfigure[][]{
\includegraphics[scale=1]{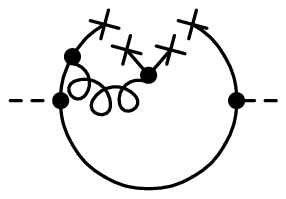}
\label{fig:4q_quark_prop_sub1}
}
\subfigure[]{
\includegraphics[scale=1]{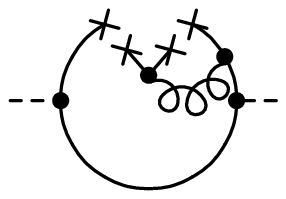}
\label{fig:4q_quark_prop_sub2}
}
\subfigure[]{
\includegraphics[scale=1]{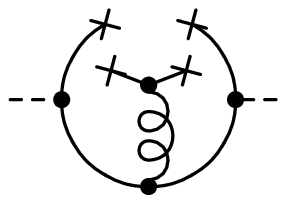}
\label{fig:4q_quark_prop_sub3}
}
\caption{Soft gluon momenta four quark condensate contributions from higher order quark propagators and quark field expansion.}
\label{fig:4q_quark_prop}
\end{figure}

\begin{figure}
\centering
\subfigure[]{
\includegraphics[scale=1]{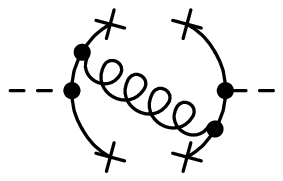}
\label{fig:4q_gluon_prop_sub1}
}
\subfigure[]{
\includegraphics{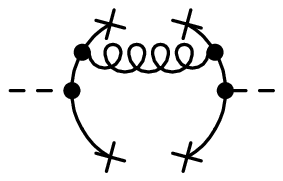}
\label{fig:4q_gluon_prop_sub2}
}
\caption{Hard gluon momenta four quark condensate contributions from higher orders of the perturbative expansion.}
\label{fig:4q_gluon_prop}
\end{figure}

We now turn our attention to four-quark contributions in \eqref{eq:def_corr_2_flavor}.
The typical four-quark condensates for the $\alpha_s$ corrections are of the form
\begin{equation} \label{eq:4q_gen}
	\langle \bar \psi \Gamma \tau \psi \bar \psi \Gamma^\prime \tau^\prime \psi \rangle {\rm L} \: ,
\end{equation}
where $\Gamma \in \left\{ \mathds{1},\gamma_\mu, \sigma_{\mu\nu}, \gamma_5 \gamma_\mu, \gamma_5 \right\}$ is a Dirac matrix and $\tau \in {\rm U}(N_{\rm f}) \otimes {\rm U}(N_{\rm c})$ is a matrix in flavor and color space (respectively, $N_{\rm c,f}$ is the number of flavors and colors).
$L$ denotes an n-fold tensor in Minkowski space constructed by suitable combinations or contractions of $\left\{ v_\mu, g_{\mu\nu}, \epsilon_{\mu\nu\alpha\beta} \right\}$ such that \eqref{eq:4q_gen} is a Lorentz scalar, in particular even w.r.t.\ parity.
$\epsilon_{\mu\nu\alpha\beta}$ is the Levi-Civita symbol, a pseudotensor.
Their diagrammatical representation can be found in Fig.\ \ref{fig:4q_quark_prop} for the contributions with soft gluon momenta and in Fig.\ \ref{fig:4q_gluon_prop} for contributions of hard gluon momenta.

These can be grouped into such ones
which are already present in the vacuum
sum rule and others which are specific
for the medium.
For the medium specific condensates the medium velocity $v$ enters $L$.
It should be emphasized that four-quark
condensates (cf. \cite{Thomas:2007gx,Thomas:2007es,Thomas:2006nk,Ronny,Thomas:2005dc} for a
complete classification) enter in
different combinations the light
vector meson (cf. \cite{Ronny,Thomas:2005dc}) or
the nucleon sum rule (cf. \cite{Ronny,Thomas:2007gx}).

Soft gluon momentum contributions arise from non-local quark fields or higher order quark propagators.
Expanding non-local quark fields up to third order in the covariant derivative and using the equations of motion for the gluon field strength tensor gives rise to the diagrams depicted in Figs.\ \ref{fig:4q_quark_prop_sub1} and \ref{fig:4q_quark_prop_sub2}.
Inserting quark propagators with a next-to-leading order gluon line attached and using the equations of motion for the gluon field strength tensor leads to the diagram shown in Fig.\ \ref{fig:4q_quark_prop_sub3}.
Soft gluon contributions in chiral partner sum rules are proportional to the medium velocity $v$ and are therefore absent in vacuum.

Contributions with hard gluons arise from the ${\cal O}(\alpha_s)$ perturbative correction.
There are vacuum as well as medium condensates.

\section{Summary}

In summary, we try to extend the chiral partner QCD sum rules derived in \cite{Hilger:2011cq} in zero order of $\alpha_s$.
Higher order of $\alpha_s$ are related to higher-dimensional condensates.
Most notable the four-quark condensates, which are of mass dimension 6, enter such difference sum rules.
Abandoning the vacuum saturation hypothesis introduces a large number of different vacuum and medium four-quark condensates, but it is necessary to keep a clean separation of chirally odd and even terms.

Furthermore, gluon condensates of the form $\langle G^3 \rangle$ are of mass dimension 6 as well.
Possible infrared divergences linked to them and their cancellation for finite temperatures or densities have to be investigated carefully.


\end{document}